\documentstyle[aps,prb,psfig,float]{revtex}
\begin{document}
\twocolumn[\hsize\textwidth\columnwidth\hsize\csname@twocolumnfalse\endcsname
\title{$N$-representability and density--functional
construction in
curvilinear coordinates}
\author{L. De Santis$^{1,2}$ and R. Resta$^{1,3}$}
\address{$^1$INFM -- Istituto Nazionale di Fisica della Materia \\
$^2$SISSA -- Scuola Internazionale Superiore di Stud\^\i\
Avanzati, Via Beirut 4, 34014 Trieste, Italy \\ 
$^3$Dipartimento di Fisica Teorica, Universit\`a di Trieste, Strada 
Costiera 11, 34014 Trieste, Italy}
\date{December 1997}
\maketitle

\begin{abstract} In practical implementations of density--functional
theory, the only term where an orbital description is needed is the
kinetic one. Even this term in principle depends on the density only,
but its explicit form is unknown. We provide a novel solution of the
$N$-representability problem for an extended system, which implies an
explicit form for the Kohn--Sham kinetic energy in terms of the
density. Our approach is based on a periodic coordinate mapping,
uniquely defined by the Fourier coefficients of the metric.  The
density functional is thus expressed as an explicit functional of the
metric tensor: since $N$-representability is enforced, our
constructive recipe provides a variational approximation. Furthermore,
we show that our geometric viewpoint is quite naturally related to
the electron localization function (ELF), which provides a very
informative analysis of the electron distribution. Studies of ELF, as
obtained from accurate Kohn--Sham orbitals in real materials, allow
an appraisal of the variational approximate density functional. We
show that the value of an approximate functional---either the present
geometric--based one or some previous ones based on different
constructive recipes---strongly depends on the nature of the chemical
bonding in the material.  \end{abstract} 

\bigskip\bigskip

]
\narrowtext

\section{Introduction}

The celebrated basic tenet of density--functional theory
(DFT)~\cite{HK,KS,DFT,Dreizler} states that an exact description of a
many--electron system is in principle possible in terms of a single
scalar field, namely the electron density $n({\bf r})$. The
$N$-electron wavefunction contains instead redundant information; in
extendend systems, it does not even have a well defined thermodynamic
limit. However, the Hohenberg--Kohn theorem~\cite{HK} (upon which DFT
is based) does not provide a constructive scheme: for any given $N$,
the {\it exact} functional is indeed accessible only through the
many--body wavefunction.  

The enormous success of DFT resides in {\it approximate} schemes
which are constructive, and do not make explicit recourse to the
many--body wavefunction. These schemes---implemented using
first--principles ingredients---have proved over the years their
astonishingly accurate predictive power for many physical properties,
in many different materials. In all these schemes the only
wavefunction needed is a wavefunction of noninteracting electrons,
which is uniquely defined by the manifold of the occupied Kohn--Sham
(KS) single--particle orbitals,~\cite{KS,DFT,Dreizler} or
equivalently by the KS one--body reduced density matrix. The
eponymous density functional $F[n]$, Eq.~(\ref{F}) below, is the sum
of a few terms: all of them but one are almost invariably
approximated as explicit functionals of the density.  The only term
where the KS orbitals (or the density matrix) are actually needed is
$T_{\rm s}$, the kinetic energy of the noninteracting system, which
is a functional of the density in an implicit way. This qualitative
difference is of course responsible for most of the computer workload
in practical calculations, hindering amongst other things the linear
scaling of computations with the size of the system. There is
therefore a quest for approximate (though accurate enough)
expressions for $T_{\rm s}$ as explicit functionals of the density.

Historically, the first approximate form for the kinetic energy of a
system of noninteracting electrons in terms of their density is the
Thomas--Fermi (TF) one,~\cite{TF} which predates DFT by several
decades and is a very crude one; it is however exact for an extended
system of free electrons. We focus here on a different class of
approximations, which are---at variance with TF---variational: this
feature is intimately linked to the problem of $N$-representability.
In fact, whenever the approximate $T_{\rm s}$ coincides with the true
kinetic energy of an arbitrary independent--electron wavefunction,
then the variational theorem ensures that it must be no smaller that
the exact $T_{\rm s}$ of the given system. The requisite of
$N$-representability is equivalent to requiring that $T_{\rm s}$
obtains from a density matrix which is idempotent. There is clearly
an infinity of idempotent density matrices, all yielding the same
given density: amongst these infinite solutions of the
$N$-representability problem, one searches for the one having the
lowest $T_{\rm s}$ at the given density.  

In order to provide an explicit approximate (and variational)
expression for $T_{\rm s}$ one has, first of all, to provide a
constructive recipe which, starting from a given density, produces an
idempotent density matrix.  Explicit solutions of this problem have
been provided by several authors in the literature, amongst whom we
only quote Harriman,~\cite{Harriman81} Zumbach and
Maschke,~\cite{Zumbach83} and Lude\~na and
coworkers.\cite{Ludena83,Ludena,Lopez97} The present paper may thus be
considered as a continuation and a generalization of this earlier work,
where the elements of novelty are basically the following.  (i) At
variance with previous work, we are interested in extended systems. We
therefore solve the $N$-representability problem for a system of $N$
noninteracting electrons in a box of volume $V$, where periodic
(Born-von K\`arm\`an) conditions are assumed at the boundary.  Our
approximate solution coincides with the exact one for the electron gas.
(ii) Our construction uses in an essential way a periodic coordinate
mapping, in the same spirit as the one advocated by
Gygi~\cite{Gygi92,Gygi93} in electronic structure calculations.  This
provides an elegant and symmetric treatment: some of the results
obtained in the previous literature assume here an interesting
geometric meaning. The electronic energy is variationally expressed in
terms of the metric tensor as the independent variable. (iii) Our
geometric approach naturally partitions $T_{\rm s}$ into the sum of two
terms: very roughly speaking ``bosonic'' and ``Pauli''.  We show that
these two terms coincide with the volume integrals of two local
functions which are used in the literature as the main ingredients of
the electron localization function~\cite{Becke90,Savin92} (ELF). 

Our solution of the $N$-representability problem---as well as the
explicit approximate density functional based upon such solution---is
therefore fundamentally linked in a very natural way to the ELF
concept.  Since ELF provides a very informative analysis of the
electron distribution, the published results for real
materials~\cite{Savin91,Kohout93,Silvi94,Savin97}---obtained {\it
a-posteriori} from accurate density matrices---help understanding what
is good and what is bad in the approximate forms of $T_{\rm s}$.  We
find that the quality of the approximation provided by our constructive
recipe strongly depends on the kind of bonding involved in the
many--electron system. We give evidence that both the Zumbach--Maschke
recipe~\cite{Zumbach83} and our own one provide a reasonable $T_{\rm
s}$ for simple metals in the pseudopotential approximation, while they
significantly overestimate $T_{\rm s}$ whenever covalent bonding is
present. The possible directions for improvement on this point are
sketched.

The present paper is organized as follows. In Sec. II we present our
novel solution to the $N$-representability problem, essentially based
on a coordinate mapping, which transforms a reference uniform system
into the actual nonuniform one. In Sec.  III we apply such solution
to the construction of a functional which is a variational
approximation to the exact one: this functional is an explicit
functional of the metric tensor. In Sec. IV we review the
foundamental properties of ELF, a powerful tool used in the
quantum--chemistry community to analyze the electron distribution in
various real systems: we show that ELF is intimately related to
basic features of the functional. Finally, in Sec. V we outline our
conclusions and perspectives, based on some ELF analyses for real
materials.

\section{Curvilinear coordinates and $N$-representability}

For a system of independent electrons in a closed--shell configuration
the wavefunction is a single determinant: knowledge of the
one--particle reduced density matrix is equivalent to a complete
knowledge of the wavefunction. The spin--integrated matrix $\rho({\bf
r,r'})$ is twice a projector, which indeed projects over the (doubly
occupied) one--particle orbitals. We consider a system of $N$
electrons in a box of volume $V$, obeying periodic boundary
conditions. The average density is $n_0 = N/V$, the density is $n({\bf
r}) = \rho({\bf r,r})$, and the idempotency condition is written:
\begin{equation} \int_{V} \! d {\bf r''} \, \rho({\bf r,r''})
\rho({\bf r'',r'}) = 2 \, \rho({\bf r,r'}) .  \label{idem}
\end{equation}

We start with a homogeneous system of $N$ non-interacting electrons
at the same density $n_0$, for which we use $\mbox{\boldmath{$\xi$}}$
as a space coordinate.  For this system the canonical orbitals are,
by symmetry, the plane waves ${\rm e}^{i {\bf k}_l
\mbox{\boldmath{$\xi$}}}/\sqrt{V}$, where ${\bf k}_l$ are the
reciprocal vectors determined by the boundary conditions. By choosing
to occupy the $N/2$ orbitals of lowest energy, the density matrix is:
\begin{equation}
\rho_0(\mbox{\boldmath{$\xi$}},\mbox{\boldmath{$\xi$}} ') = \frac 2V
\sum_{l=1}^{N/2} {\rm e}^ {i{\bf k}_l (\mbox{\boldmath{$\xi$}} -
\mbox{\boldmath{$\xi$}}')}, \label{densz} \end{equation} which is
obviously idempotent and yields a constant density. In the
thermodynamic limit ($N \rightarrow \infty$ and $V \rightarrow
\infty$ at constant $n_0$)  the ${\bf k}_l$ set becomes dense.
Occupying the {\bf k}-vectors within the Fermi sphere ($|{\bf k}| <
k_{\rm F}$) Eq.~(\ref{densz}) yields the well known electron--gas
result: \begin{equation}
\rho_0(\mbox{\boldmath{$\xi$}},\mbox{\boldmath{$\xi$}} ') = n_0
\frac{ 3 j_1(k_{\rm F} | \mbox{\boldmath{$\xi$}}
-\mbox{\boldmath{$\xi$}}'|)}{k_{\rm F} | \mbox{\boldmath{$\xi$}}
-\mbox{\boldmath{$\xi$}}'|} .  \label{egas} \end{equation}

At this point we introduce a generic curvilinear coordinate precisely
of the same kind as introduced by Gygi in the field of
electronic--structure calculations.~\cite{Gygi92,Gygi93} We therefore
define a twice differentiable invertible map $\mbox{\boldmath{$\xi$}}
\rightarrow {\bf r}(\mbox{\boldmath{$\xi$}})$, periodic over $V$, whose
Riemannian metric tensor is: \begin{equation} \label{tensor} g_{ij} =
\frac{\partial r^k}{\partial \xi^i}\frac{\partial r^k}{\partial \xi^j} 
. \end{equation} Summation over repeated indices is understood
throughout. A generic plane wave of momentum {\bf k} is transformed as:
\begin{equation} \label{pw} \frac 1{\sqrt V} {\rm e}^{i {\bf k}
\mbox{\boldmath{$\xi$}}} \longrightarrow \chi_{\bf k}({\bf r}) = \frac
1{\sqrt V} g^{-{1 \over4}}({\bf r}) {\rm e}^{ i {\bf k}
\mbox{\boldmath{$\xi$}}({\bf r})} , \end{equation} where $g = \det \{
g_{ij} \}$, and $g^{-\frac12}$ is the Jacobian $| \partial
\mbox{\boldmath{$\xi$}}/\partial {\bf r}|$  of the inverse
transformation. Notice that the orbitals  $\chi_{\bf k}({\bf r})$ have
a {\bf k}-independent density, and are therefore ``equidensity
orbitals'' in Harriman's~\cite{Harriman81} nomenclature.  The density
matrix in the new coordinates is: \begin{eqnarray} \label{dens}
\rho({\bf r,r'}) & = & \rho_0(\mbox{\boldmath{$\xi$}}({\bf
r}),\mbox{\boldmath{$\xi$}}({\bf r'})) = \nonumber \\ & = & \frac 2V
g^{-\frac14}({\bf r}) g^{-\frac14}({\bf r'}) \sum_{l=1}^{N/2} {\rm e}^
{i{\bf k}_l [ \mbox{\boldmath{$\xi$}}({\bf
r})-\mbox{\boldmath{$\xi$}}({\bf r'})]} . \end{eqnarray} The
corresponding transformed density is \begin{equation}\label{denst}
n({\bf r})=n_0 \, g ^{-\frac12}({\bf r}). \end{equation} In the novel
coordinates we thus have a nonhomogeneous system, with the same average
density as the homogeneous one, and whose density matrix is idempotent
by construction.

We are now ready to attribute physical content to the above
mathemathics. Suppose that the density $n({\bf r})$ of an electronic
system is given. Then we may look for a coordinate transformation
$\mbox{\boldmath{$\xi$}} \longrightarrow {\bf
r}(\mbox{\boldmath{$\xi$}})$ which maps the uniform density into the
given density: a necessary and sufficient condition is
Eq.~(\ref{denst}). The solution is nonunique, since several different
maps share the same Jacobian $g ^{-{1\over2}}$: we will argue below
about an optimal solution, using a variationally adaptive metric in
the sense of Gygi.~\cite{Gygi92,Gygi93} Replacement into
Eq.~(\ref{dens}) yields the explicit form: \begin{equation} \rho({\bf
r,r'}) = \frac 2N n^{\frac12}({\bf r}) n^{\frac12}({\bf r'})
\sum_{l=1}^{N/2} {\rm e}^ {i{\bf k}_l [ \mbox{\boldmath{$\xi$}}({\bf
r})-\mbox{\boldmath{$\xi$}}({\bf r'})]} . \label{dens2} \end{equation}

In one dimension the solution of Eq.~(\ref{denst}) is unique, and we
get here the periodic analogue of the Harriman
construction.\cite{Harriman81} In three dimensions, our result is
related to the work of Zumbach and Maschke,~\cite{Zumbach83} the
differences being that we deal with periodic systems, and we provide a
more general explicit construction. A coordinate mapping, similar in
spirit to the present one (and called ``local scaling
transformation''), has been previously introduced by  Lude\~na and
coworkers for spherical atoms.\cite{Ludena}

\section{Density functional}

The energy of the electronic system in the external potential
$v_{\rm ext}$ is written, within DFT, as:\cite{DFT,Dreizler}
\begin{eqnarray} E[n] & = & \int_{V} \! d {\bf r} \, n({\bf r})
v_{\rm ext}({\bf r}) + F[n] ; \label{total} \\ F[n] & = & T_{\rm
s}[n] + \frac12 \int_{V} \! d {\bf r} \int \! d {\bf r'} \,
\frac{n({\bf r})n({\bf r'})}{|{\bf r - r'}|} + E_{\rm xc}[n] ,
\label{F} \end{eqnarray} where atomic Hartree units have been used.
As already anticipated, basically all the available constructive
approximations to DFT provide  $E_{\rm xc}$ as an explicit
functional of the density $n({\bf r})$, while instead the kinetic
energy term $T_{\rm s}$ is: \begin{equation} T_{\rm s} = \frac12
\left. \int_{V}  \! d {\bf r} \, \nabla_{\bf r} \nabla_{\bf r'}
\rho({\bf r,r'}) \right|_{\bf r = r'}. \label{kine1} \end{equation}
The ground electronic energy, Eq.~(\ref{total}), is therefore an
explicit functional of the density matrix, which has to be minimized
under the constraints of idempotency, Eq.~(\ref{idem}), and electron
number.  

Replacement of our ansatz density matrix, Eq.~(\ref{dens2}), in the
above expressions provides an upper bound to the electronic energy,
explicitly expressed solely in terms of the density and of the
metric.  The approximate kinetic energy is: \begin{equation}
\tilde{T}_{\rm s} = \sum_{l=1}^{N/2} \int_{V}  \! d {\bf r} \,
|\nabla \chi_{ {\bf k}_l}({\bf r}) |^2 . \end{equation}
Using then Eq.~(7) of Ref.~\onlinecite{Gygi93}, the expectation value
of the kinetic energy over a $\chi_{\bf k}$ orbital is the sum of two
positive terms: \begin{equation} \langle \chi_{\bf k} | T | \chi_{\bf
k} \rangle = \frac{ k_i k_j}{2V} \int_{V}  \! d
\mbox{\boldmath{$\xi$}} \, g^{ij} + \frac{1}{2V} \int_{V}  \! d
\mbox{\boldmath{$\xi$}} \,  \, A_i g^{ij} A_j  , \end{equation} where
the ``gauge potential'' is: \begin{equation} \label{gauge} A_i =
\frac14 \frac {\partial \ln g}{\partial \xi^i}.  \end{equation} After
summing over the $N/2$ doubly occupied states, we get
\begin{equation} \tilde{T}_{\rm s}[n] = \tilde{T}_{\rm P}[n] + T_{\rm
B}[n] , \end{equation} where the reason for the notations will be
clear in a moment. 

Using Eq.~(\ref{denst}), we cast the gauge term as: \begin{equation}
T_{\rm B}[n] = \frac{n_0}{32} \int_{V} \! d {\bf r} \,
g^{-\frac{5}{2}}({\bf r}) \, | \nabla g({\bf r}) |^2 = \frac18
\int_{V} \! d {\bf r} \, \frac{| \nabla n({\bf r}) |^2}{n({\bf r})}
.  \end{equation} In the latter expression, we notice that the
metric formally disappeares from the gauge term, which is indeed
identical to the so--called von Weizs\"acker energy
functional.\cite{Dreizler} This energy coincides with the kinetic
energy of a system of noninteracting bosons in their ground state,
having the given density $n({\bf r})$: with this specific meaning,
we may refer to $T_{\rm B}$ as to the ``bosonic'' energy. It is easy
to prove that $T_{\rm B}$ is a lower bound to the kinetic energy
$T_{\rm s}$ of a system of noninteracting fermions,\cite{Dreizler}
and coincides with $T_{\rm s}$ only in the trivial case $N\!=\!2$,
where the Pauli principle has no effect (we are considering singlets
only). In all the interesting cases, there is an excess kinetic
energy $T_{\rm P}$ due to the Pauli principle: \begin{equation}
T_{\rm P} = \frac12 \left. \int_{V}  \! d {\bf r} \, \nabla_{\bf r}
\nabla_{\bf r'} \rho({\bf r,r'}) \right|_{\bf r = r'} \!\! - \frac18
\int_{V} \! d {\bf r} \, \frac{| \nabla n({\bf r}) |^2}{n({\bf r})}.
\label{Tp} \end{equation} Our expression for  $\tilde{T}_{\rm P}[n]$
as a function of the metric shall therefore be a variational
approximation to the true excess Pauli energy $ T_{\rm P}$.  

The sum over the occupied states in $\tilde{T}_{\rm P}$ is most
easily evaluated if we assume a cubic box. If we define $E_0$ as the
kinetic energy of the homogeneous system: \begin{equation}
\label{energy} E_0 = \sum_{l=1}^{N/2} |{\bf k}_l|^2, \end{equation}
it is then easy to recast $\tilde{T}_{\rm P}[n]$ as: \begin{equation}
\label{pauli} \tilde{T}_{\rm P}[n] = \frac {E_0}{3 V} \int_V \! d{\bf
r} \; g^{-\frac12} \, \mbox{tr} \{g^{ij}\} .  \end{equation} We
further notice that in the thermodynamic limit one has:
\begin{equation} E_0 = \frac{3}{10} N k_{\rm F}^2 =  c_{\rm F} V
n_0^{\frac53}, \label{egas2} \end{equation} where  $c_{\rm F} =
\frac{3}{10} (3 \pi^2)^{\frac23}$.  

Putting all the previous formulas together and approximating $T_{\rm
s}$ in Eq.~(\ref{F}) with $\tilde{T}_{\rm s}$, we obtain an
approximate $F[n]$ as an explicit functional of the density and of
the metric tensor. Since the density---owing to
Eq.~(\ref{denst})---is in turn a function of the metric tensor, we
use the latter as the independent variable. Eventually, the
electronic energy of the system, Eq.~(\ref{total}), is a variational
explicit functional of the metric tensor $g^{ij}({\bf r})$.  This
functional can be regarded as the periodic analogue of the one of
Zumbach and Maschke,~\cite{Zumbach83} expressed in more compact form
in terms of a different variable. Furthermore the explicit occurrence
of the periodic metric in the $\tilde{T}_{\rm P}$ term makes feasible
an adaptive optimization of the metric. Upon closely following Gygi's
approach,\cite{Gygi93} the Fourier coefficients of the periodic
metric are the natural variational parameters of the problem.

Finally, we end this section just noting that in the trivial case
$v_{\rm ext}({\bf r}) = 0$ all of the kinetic energy is due to
$T_{\rm P}$, since the density is constant and $T_{\rm B}$ vanishes. 
Furthermore the metric is the identity and the approximate kinetic
energy equals the exact one: $ \tilde{T}_{\rm s} = T_{\rm s} = E_0$,
Eq.~(\ref{egas2}). The approximate functional $F[n]$ coincides with
the exact one, {\it including} its exchange--correlation term if the
exact electron--gas data\cite{XC} are used therein (as usual). This
suggests that the approximate functional should work reasonably well
for a system close enough to the electron gas, such as a simple metal
within a pseudopotential scheme.

\section{Electron localization function}

The kinetic energy $T_{\rm s}$ can be thought of as the integral over
$V$ of a kinetic energy density $\tau({\bf r})$. It is well known
that the expression for $\tau({\bf r})$ is not unique: we use the
form suggested by Eq.~(\ref{kine1}), namely, \begin{equation}
\tau({\bf r}) = \left. \frac12 \nabla_{\bf r} \nabla_{\bf r'}
\rho({\bf r,r'}) \right|_{\bf r = r'} \label{kine2} , \end{equation}
which is everywhere positive, and coincides with the choice made in
the ELF literature.\cite{Savin92} By analogy, one defines the Pauli
excess energy density, after Eq.~(\ref{Tp}), as: \begin{equation}
\tau_{\rm P}({\bf r}) = \left.  \frac12 \nabla_{\bf r} \nabla_{\bf
r'} \rho({\bf r,r'}) \right|_{\bf r = r'} \!\! - \frac18 \frac{|
\nabla n({\bf r}) |^2}{n({\bf r})} \label{pauli2} .  \end{equation}
This function is the main ingredient of ELF,\cite{Becke90} in the
formulation due to Savin {\it et al.},\cite{Savin92} who write the
function as: \begin{equation} {\cal E}({\bf r}) = \left\{ 1 + \left[
\frac{\tau_{\rm P}({\bf r})}{c_{\rm F} \, n^{\frac53}({\bf r})}
\right]^2 \right\}^{-1} \label{elf} .  \end{equation} This function
by design takes values between zero and one: several of its features
are remarkable. In the homogeneous electron gas, owing to
Eq.~(\ref{egas2}), the ELF equals 1/2 at any density.  In a
nonhomogenous system ${\cal E}({\bf r})$ assumes values close to its
upper bound 1 in the regions of space where there is an high
probability of finding a pair of electrons with antiparallel spins
(or an isolated electron):\cite{Becke90,Savin92} with this meaning,
we may say that ${\cal E}({\bf r})$ close to 1 characterizes space
regions where the electron distribution is ``bosonic''. Conversely, 
${\cal E}({\bf r})$ is close to 0 in low--density regions.\cite{nota}

\begin{figure}
\centerline{\psfig{figure=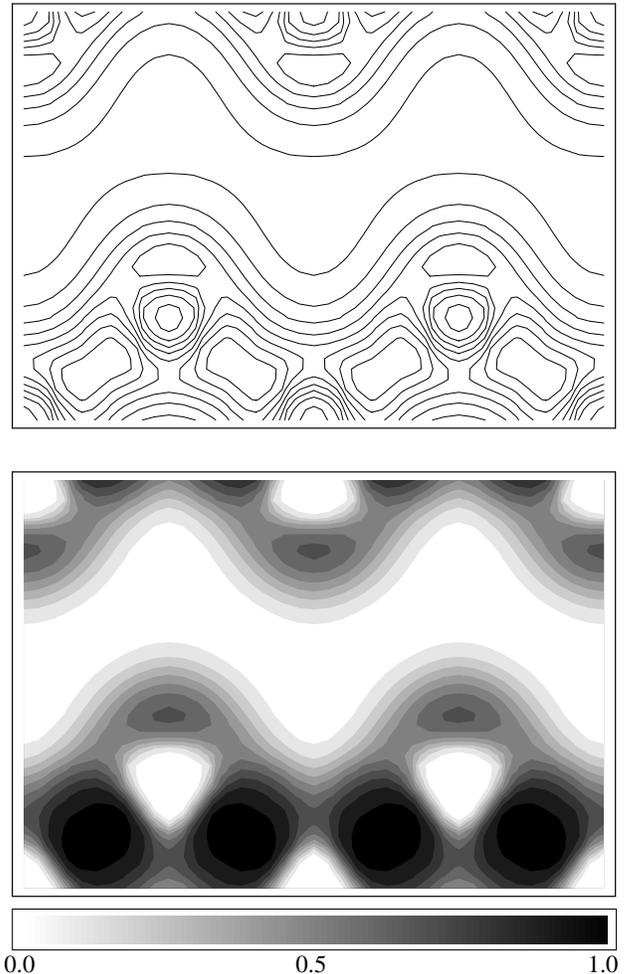,width=8.5cm}}
\caption {Pseudocharge density contour plot (upper
panel) and the corresponding ELF (lower panel) for bulk silicon in the
$[110]$ cristalline plane. The grey--scale is also shown: dark regions
correspond to large ELF values. The maximum ELF value at the bond
center is 0.96.} \label{f:Fig1} \end{figure}

\begin{figure} 
\centerline{\psfig{figure=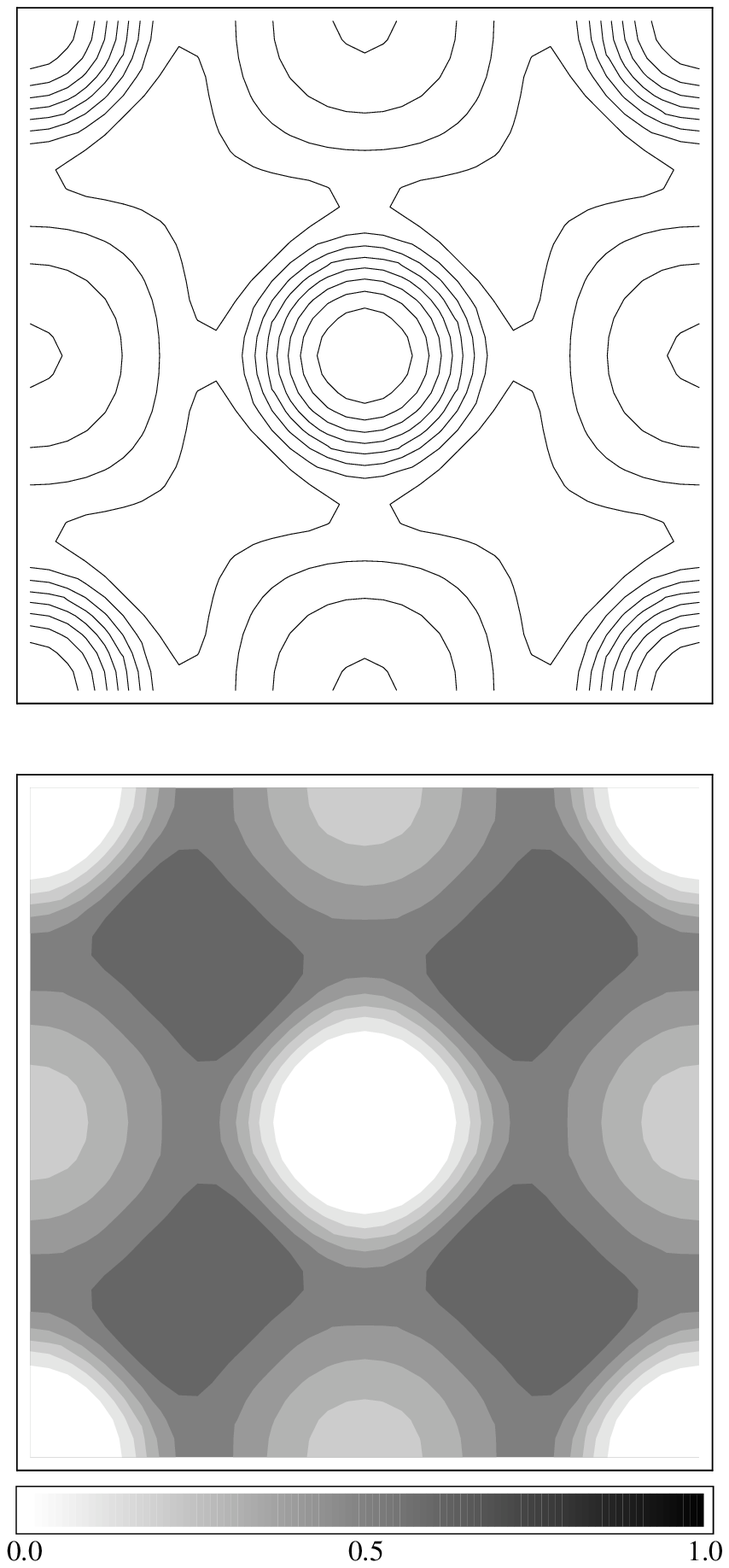,width=8.5cm}}
\caption {Pseudocharge density  contour plot (upper
panel) and the corresponding ELF (lower panel) for bulk aluminum in
the $[100]$ cristalline plane. The maximum ELF value in the interatomic
region is 0.61.} \label{f:Fig2} \end{figure}

Detailed ELF analyses have been performed for many systems of
chemical interest.\cite{Savin91,Kohout93,Silvi94} For heavy atoms,
the ELF perspicuously localizes in space the different electronic
shells: furthermore, when the valence shell is considered, the ELF
provides a very meaningful description of the chemical bond for many
classes of compounds.\cite{Savin92,Savin97} 

For the sake of simplicity, we consider only the class of $sp$-bonded
materials which are accurately described within a norm--conserving
pseudopotential scheme:\cite{Pickett89} this class includes simple
metals, covalent semiconductors, simple ionic solids, and many other
disparate materials. We have in this case by construction only a
single shell (the valence $sp$ one of each atom involved), and the
ELF allows a very meaningful analysis of the chemical bonding.
Perhaps the most spectacular performance is the ability to
perspicuously distinguish in a very clearcut way between metallic
bonding and covalent bonding.  This is illustrated in the lower
panels of Figs.~(\ref{f:Fig1}) and (\ref{f:Fig2}), where we plot the
function ${\cal E}({\bf r})$ for two paradigmatic crystalline
materials: respectively, silicon and aluminum.

In the covalently bonded system of Fig.~\ref{f:Fig1}, the bond electron
pairs forming the typical ``zig-zag'' chain in the  $[110]$ plane are
clearly visible. The very dark regions in the lower panel indicate in
fact the strong  bosonic character of the charge density in the bond
region.  Actually, ${\cal E}({\bf r})$ attains the maximum value
of 0.96 at the bond center, thus indicating an extremely strong pairing
between opposite--spin electrons.

A completely different picture emerges for our simple metallic system,
Fig~\ref{f:Fig2}. The ELF plot in the lower panel shows---outside the
core regions---a large grey area, which correspond to a jellium--like
(or Thomas--Fermi) ELF value.  Actually, the maximum value attained by
${\cal E}({\bf r})$ is only 0.61. This is agreement with the usual
picture of a simple metal, where the (pseudo) electrons are excluded
from the core region, while behaving essentially as free particles in
the rest of the material.\cite{pseudo}

Comparison of the two ELF plots provides therefore the most significant
and perspicuous visualization of the important {\it qualitative}
difference between the covalent bond and the metallic one. In the two
classes of materials the Pauli principle plays quite a different role. 
At variance with the ELF, the corresponding  charge density plots,
(upper panels in Figs~\ref{f:Fig1} and \ref{f:Fig2}) are much less
informative, and do not qualitatively discriminate between the two
different kinds of chemical bonds.

\section{Conclusions and perspectives}

The experience gained in investigating the ELF in real materials
helps understanding the meaning and the limits of the approximate
density functional such as the Zumbach--Maschke\cite{Zumbach83} one,
as well as of the generalization proposed here in Sec. III. 

Our explicit ansatz of Eq.~(\ref{dens2}) leads to the Pauli excess
energy $\tilde{T}_{\rm P}$ of Eq.~(\ref{pauli}). It is interesting to
see the consequences for the ELF, since the ansatz clearly leads to
replacing the Pauli excess energy density, Eq.~(\ref{pauli2}), with
\begin{equation} \tilde{\tau}_{\rm P}({\bf r}) = \frac {E_0}{3 V} \,
g^{-\frac12} \, \mbox{tr} \{g^{ij}\}  = \frac13 c_{\rm F}
n_0^{\frac53} g^{-\frac12} \, \mbox{tr} \{g^{ij}\}, \label{pauli3}
\end{equation} where the thermodynamic limit, Eq.~(\ref{egas2}), has
been used.  Considering now the inequality \begin{equation} \frac13
\mbox{tr} \{g^{ij}\} \geq  [ \mbox{det} \{g^{ij}\} ]^{\frac13} =
g^{-\frac{1}{3}} = \frac{n^{\frac23}({\bf r})}{n_0^{\frac23}},
\end{equation}  we get for the approximate Pauli energy density the
lower bound: \begin{equation} \tilde{\tau}_{\rm P}({\bf r}) \geq
c_{\rm F} \, n^{\frac53}({\bf r}) . \label{pauli4} \end{equation}
Comparing with the ELF definition, Eq.~(\ref{elf}), one easily
realizes that even the optimal choice of the metric tensor will
unavoidably provide values of  ${\cal E}({\bf r})$ which are smaller
than 1/2 everywhere. 

In this work we have shown that a fundamental link exists between
$N$-representability, approximate explicit functionals, and ELF. 
This link emerges very naturally within the geometric approach upon
which our work is based. The same geometric approach, however, also
indicates very clearly the limits of the approximate form of the
kinetic energy for real materials which we have found here. In fact the
final considerations of the previous paragraph imply that our
constructive recipe, as well as the previous one of Zumbach and
Maschke,\cite{Zumbach83} are a good approximation only for systems
where the bonding is metallic, while it necessarily overestimates the
kinetic energy (and the total energy) whenever covalent bonding is
present.  Looking more closely, this major limitation owes to the
occurrence of equidensity orbitals in our ansatz density matrix,
Eq.~(\ref{dens}), which occurrence can be further traced back to the
choice of the uniform electron gas as the reference system upon which
we perform the coordinate transformation. 

This naturally suggests the directions for improvements: one should
start from a reference model system other than the uniform electron
gas, having instead some covalent bonding features already built in. 
Interestingly, the use of a model reference system designed to
reproduce---after coordinate mapping---some desirable features of the
real one has been proposed in the most recent work of  Lude\~na and
coworkers.\cite{Ludena,Lopez97} These authors, however, focus on a
spherical atom having several electronic shells: here instead we are
discussing a condensed system with only one valence shell, within a
pseudopotential scheme.

\section*{Acknowledgments}

R.R. is grateful to A. Savin for several discussions about ELF.


\end{document}